# Hybrid Design Tools—Image Quality Assessment of a Digitally Augmented Blackboard Integrated System

**Ovidiu Banias [1],\* and Camil Octavian Milincu [2]**

[1] Automation and Applied Informatics Department, Faculty of Automation and Computer Science, Politehnica University of Timisoara, 300223 Timisoara, Romania; ovidiu.banias@aut.upt.ro

[2] Architecture Department, Faculty of Architecture and Urban Planning, Politehnica University of Timisoara, 300223 Timisoara, Romania; milincucamil@yahoo.com

**\*** Correspondence: ovidiu.banias@aut.upt.ro; Tel.: +40-732-673-387

**Abstract:** In the last two decades, Interactive White Boards (IWBs) have been widely available as a pedagogic tool. The usability of these boards for teaching disciplines where complex drawings are needed, we consider debatable in multiple regards. In a previous study, we proposed an alternative to the IWBs as a blackboard augmented with a minimum of necessary digital elements. The current study continues our previous research on hybrid design tools, analyzing the limitations of the developed hybrid system regarding the perceived quality of the images being repeatedly captured, annotated, and reprojected onto the board. We validated the hybrid system by evaluating the quality of the projected and reprojected images over a blackboard, using both objective measurements and subjective human perception in extensive and realistic case studies. Based on the results achieved in the current research, we conclude that the proposed hybrid system provides good quality support for teaching disciplines that require complex drawings and board interaction.

**Keywords:** design tools; hybrid systems; interactive whiteboard; digital; blackboard; teaching

## 1. Introduction

Interactive White Board (IWB) systems are trendy nowadays, being perceived as the logical evolutionary step after switching from blackboards to whiteboards. They have the possibility to interact within certain limits with the presented content, allowing the creation of annotations, as well as the capture and subsequent sharing of the material. Several studies [1–8] have been conducted on the use of IWB focusing mainly on the usage in the primary education. Although the results are promising in some areas, the disadvantages of the system must be highlighted, both in terms of physical components and interactiveness.

IWBs are complex systems that require high costs both in acquisition and in maintenance. IWBs in comparison with analogue systems are slow and non-intuitive [9]. With current options, it is not possible to make an interactive presentation containing complex drawings without considerable effort, at a slow pace and constantly accessing some menus. For lectures, especially when there is no alternative to an analog writing surface, it is necessary to prepare the material prior to the presentation. The interactive component is thus restricted and limited. Although the systems use simple software components, a process of adaptation and rehearsal is required from the teacher, requiring an in-depth knowledge of the technology, especially if it is desired to allow students to interact with the board during the presentations [10]. Based on our experience, the interactive capacity of the IWB is sometimes overlooked, the whiteboard surface being used for projection purposes only because, in order to prevent the premature wear of the surface, the usage of dry-erase markers is often avoided. Therefore, a traditional blackboard is used as complementary tool [11].

Another disadvantage of IWBs would be the limitations regarding the drawing line quality in terms of expressiveness [11]. Current capabilities of representation could be acceptable in some areas of teaching but are insufficient in the technical disciplines as design and architecture, where the expressiveness and ambiguity of sketches are desirable features.

As an alternative to the current IWBs, in a previous study we proposed a hybrid system consisting of a blackboard as an analogical medium augmented with a minimum of necessary digital elements [12]. The system is relevant and convenient for technical disciplines where complex drawings are needed. The major challenge the proposed hybrid system is facing, excluding the software and hardware challenges is the offering of a projected image of a good quality, good enough to be able to support the teaching process as a better alternative to current IWBs.

The current study continues our previous research on hybrid design tools analyzing the limitations of the developed hybrid system regarding the perceived quality of the images being repeatedly captured, annotated, and reprojected onto the board. We validate the hybrid system by evaluating the quality of the projected, annotated, and reprojected images over a blackboard.

The paper is structured as follows. In the next section we present the advantages and disadvantages of using whiteboards and blackboards in the context of teaching technical disciplines that require complex drawings and board interaction during the presentation. We continue with presenting the assessment methodologies we used to evaluate our previously developed hybrid system. The hardware setup is also briefly described. We continue further with four case studies developed in one of the classrooms of the Politehnica University of Timisoara, Faculty of Architecture and Urban Planning, Architecture Department, using a blackboard as a projecting surface. In the last section we discuss the results and we draw the conclusions.

## 2. Context and Related Work

IWBs are currently fashionable tools [13]. However, there are discussions about how they are being used, and more important, about the possibility to become a common element in the teaching process. For making it happen, the system should have a significant analogical component [14]. As an alternative to IWBs, in a previous study [12], a proof-of-concept system was developed to digitally augment a presentation surface. The decision to use a minimum of digital elements in combination with a blackboard surface was taken based on several considerations presented below.

Whiteboards are used with dry-erase markers, mitigating the chalk dust issues. The erasing is easier, various markers are available, but they are considered unreliable: they get dry, used and worn easily. Also, the remaining quantity of ink in the marker is difficult to estimate, so teachers often need to have around many markers. To express or emphasis an idea by drawing, lines are being drawn by two techniques: modulation of the thickness and variation of the intensity [15]. Concerning the expressiveness of the drawn lines, significant differences can be observed between the whiteboards and the blackboards, the blackboards performing much better.

A downside of the whiteboard is the glossy finishing of the working surface. The glossiness of the surface combined with the direction of lighting causes lots of distracting reflections, a phenomenon observed regardless of the position in the workroom. In the case of using a blackboard, because of its rough surface needed for chalk writing, the disturbing reflections are avoided. Although the marker generated lines have a higher contrast than the chalk, the marker line thickness is significant thinner. In the case of dry-erase markers, the contrast decreases as the ink reserve is depleted, creating the effect of a dry line which has a reduced density. To overcome this shortcoming and compensate the lack of density through a larger thickness of the line, the user must press harder on the marker and deform the tip. In the case of using chalk, the density of the line remains constant.

Although the whiteboard offers a good projection support for static presentations containing text and diagrams that do not require interaction, there are significant disadvantages regarding interaction and expressiveness: practicability of line generation, line and hatch density, elements of multisensory perception [16]. In the case of utilizing a blackboard, chalk is used to draw on sandblasted glass. The advantage of using a blackboard is the easiness of controlling the thickness and density of the line by using the same writing instrument and without using any other means except the exerted pressure. Also, different line thicknesses or hatches with various intensities on vast surfaces can be accomplished by using the edge of the chalk over a blackboard.

Drawing precision is also better when using a chalk over a blackboard than using a marker over a whiteboard. The same precision issue occurs when drawing on paper with a ballpoint pen [17]. Usually straight lines with different orientation are being drawn, and seldom it is necessary to draw more than three intersecting lines. The quality of the drawing and of the writing decreases due to the sliding of the marker tip on the whiteboard sleek surface. By comparing the same person's drawings, a better control can be noticed in the case of using the blackboard because chalk has a higher coefficient of friction [11]. The differences are noticeable mostly in the starting point of the line and where increased pressure is being applied.

Regarding multisensory perception, the tactile and audible feedback [16] has the same importance as gesture and can bring new elements into the presentation. In the case of using whiteboards these perceptions are almost unnoticeable, but tactile and auditive elements become significantly more important when using a blackboard. There is a strong correlation between the pressure and the speed of drawing given by the chalk, just like an important auditive component [11]. Tests using whiteboard markers show that an expressive, accentuated drawing with great pressure and speed can create a reversed phenomenon [11]. The capillary system of the marker is not able to provide enough ink flow and the result is different to the one expected because the lines which are needed to be thickened have a reduced density.

In a previous study [18] we analyzed the process of teaching design discipline and some shortcomings were noticed due to the students' prejudices and inefficient use of the resources at their disposal. The students were given the task of designing a piece of furniture and had full freedom in the choice of the modeling and representation techniques: manual drawing, digital drawing, computer-aided design, 3D modeling, layout and photography, as well as any hybrid technology. Contrary to the expectations, no original or advanced use of digital means by the students was observed. Employing exclusively digital means of modeling and representation produced results that lacked formal complexity compared to those designed and drafted in analogical environments.

From the perspective of design process, a discipline that requires complex drawings and interaction, hybrid techniques reduce the "ideation gap" in respect to employing exclusively digital or analogical means. Hybrid techniques also increase the processing speed or what Thomas Dorta names "design flow" [9]. The development of coherent ideas is supported by the opportunities of a quick working method, any exclusive use of a method, whether it is digital or analog, having no positive effects [19–21] during design process. Boards are suitable tools for design alternatives unlike software systems which use predefined elements, being more capable to adapt to certain situations. Using boards in presentations at the expense of digital items can shift the student's preferred way of using exclusively digital design tools and predefined functions. In addition, the ambiguous nature of hand-drawn sketches favors variation and diversity, personal discussions and interpretations.

## 3. Methodology and Hardware Setup

In the previous researches we presented hybrid design tools and developed a hybrid system [12] as a proof of concept to be able to validate our proposed solutions. In this paper, we continue further the research by validating the proposed solution through evaluation of the quality of the projected and reprojected images over a blackboard, using both objective measurements and subjective human opinion. As reprojected images are subject to distortion and blurriness for objective measurements we employed the Structural Similarity Index (SSIM) to be able to evaluate the difference between two consecutive projected and reprojected images.

The challenges the proposed hybrid system is facing are on one hand the offering of a projected image of a good quality to be able to support the teaching process and on the other hand becoming an alternative to the current IWBs. The repeated process of capture and projection in the context presented above could diminish drastically the quality of the image with respect to skewing, blurriness, and luminosity. We evaluated the quality of the projected/reprojected images objectively by calculating the Structural Similarity Index (SSIM) for two or more consecutive projected images and subjectively by asking opinion of both human experts as design teachers, and unexperienced humans as design students that are supposed to benefit from the interactive teaching. We have chosen as an image quality metric the Structural Similarity Index (SSIM) due to better results in comparing two images [22,23] than Mean Square Error (MSE) and Peak Signal to Noise Ratio (PSNR) metrics.

*3.1. Hybrid Projection System Overview*

The proposed hybrid system presented in [12] is composed of five hardware components: laptop, projector, camera, Arduino microcontroller, smart phone, and three software applications for PC, smart phone, and microcontroller. The hybrid system depicted in Figure 1 is meant for supporting the teaching process by repeating the following three steps:

- projecting course slides/images over a surface,
- capturing (electronically saving) the image of the surface containing previously projected slide/image and human interaction as drawings in chalk over a blackboard,
- reprojection of the saved images on demand.

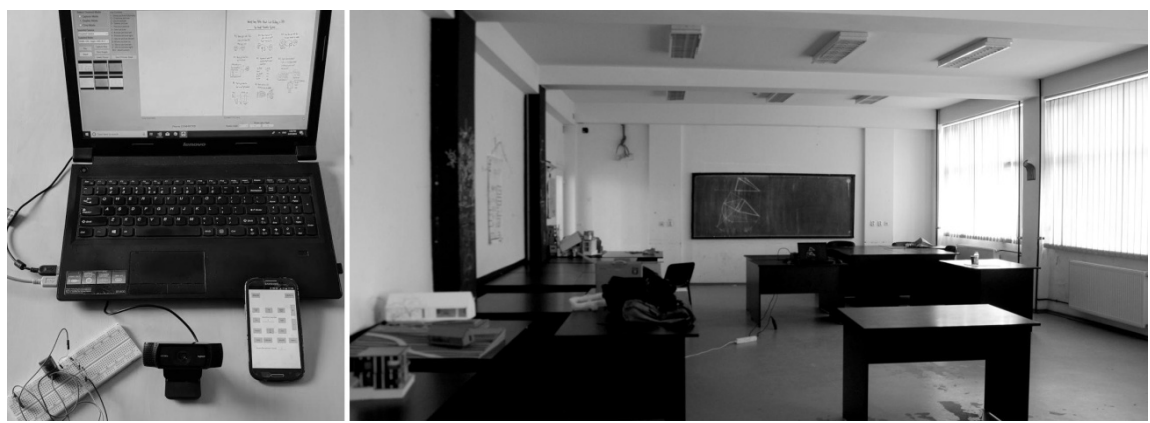

**Figure 1.** Proof of concept hybrid system setup.

Minimum System requirements:

- PC,
- Camera—high resolution web camera or photo camera,
- Projector—good color light output (brightness) minimum 2000 lumens,
- Smart phone (optional) —running Android operating system,
- Arduino Uno & Bluetooth module (optional).

*3.2. Hardware Setup*

In order to evaluate the hybrid system presented in the previous section we used the following hardware components:

- Laptop: Lenovo B590,
- Projector: Benq MX511,
- Camera & Lens: Canon EOS 60D & Canon Zoom Lens EF 100–400 mm, 1:4.5–5.6 L;

and had the following setup:

- Distance from projector to the board: 280 cm,
- Distance from camera lens to the board: 750 cm,
- Captured image dimension: 130 × 95 cm,
- Distance between image markers: 126 × 85 cm,
- Ambient/board illuminance level: 205 lux,
- Projector/board illuminance level: 820 lux,
- Height of projected Lena's image [24]: 96 cm.

The quality of the captured and projected images is affected by projector, camera, and lens quality. Quality in general could be assessed objectively and subjectively. In the context of current research, we refer to image quality from the perspective of: sharpness, luminance, and contrast for the hybrid system to be able to support the teaching process up to an acceptable visual human opinion. Visual contrast is defined as a perceptual image attribute because the assessment of the contrast of images is influenced by previous experiences and subjective factors [25,26]. The perceived contrast cannot be determined correctly by analyzing only some points in the image, thus global analysis is required. This is influenced by several factors, such as the subjective preset interest areas. The projector used for the current research could be classified as low quality, producing only 2700 lumens white light output and only 700 lumens color light output, where color light output represents the color brightness as a standard in measuring the projector's ability to reproduce a color. The quality of the projected images could be improved further as needed by using a better projector with higher color light output values. Otherwise, the camera and lens used in current research offer a good quality. We conjecture that the high quality of the projector is more important in the current research context than the high quality of the camera and lens, because the luminosity is the most important observed criterion that affects the human perception.

## 4. Case Studies

The case studies were developed in one of the classrooms of the Politehnica University of Timisoara, Faculty of Architecture and Urban Planning, Architecture Department, having the hardware setup described above and using a blackboard as projecting surface. Each of the case studies were initiated by projecting a support image as a presentation slide for supporting further interactions in white and colored chalk over the blackboard. Images of the blackboard surface were captured and reprojected as needed in order to simulate the interactive teaching process where the teacher is required to return to previous drawings for further clarifications.

For objective measurements we used *ssim* and *imshowpair* Matblab functions:

- "*ssimval = ssim(A,ref)* computes the Structural Similarity Index (SSIM) value for image A using ref as the reference image",
- "*imshowpair(A,B,'falsecolor')* creates a composite RGB image showing A and B overlaid in different color bands. Gray regions in the composite image show where the two images have the same intensities. Magenta and green regions show where the intensities are different".

For subjective measurements we questioned 29 students and 5 design teachers within the Department of Architecture. Both students and teachers were presented projected and reprojected images using the proposed hybrid system and they were asked to assess the perceived image quality on a 1 to 5 scale, where 5 is the highest perceived quality and 1 is the lowest perceived quality. The scope of the assessment was to observe how the perceived image quality drops with repeating the capturing and reprojection process. We questioned both students and teachers in order to observe if there is a significant bias due to the different relation with technology of the two groups. In the following subsections, four case studies will be presented to assess the quality of the reprojected images.

*4.1. Analyzing the Use of Different Types of Design References in Reprojected Images*

In the current case study, we analyze how the networks of points, lines, or surfaces can be used as references, and how the image quality is affected by successive capture and reprojection. In Figure 2, we depict a series of projected–captured–altered–reprojected images. We evaluate objectively the similarity of each of the images in relation to the first one by calculating the SSIM value, and in Figure 3 we present the results obtained by questioning teachers and students. In Figure 4 we present the visual dissimilarities between images in different color bands.

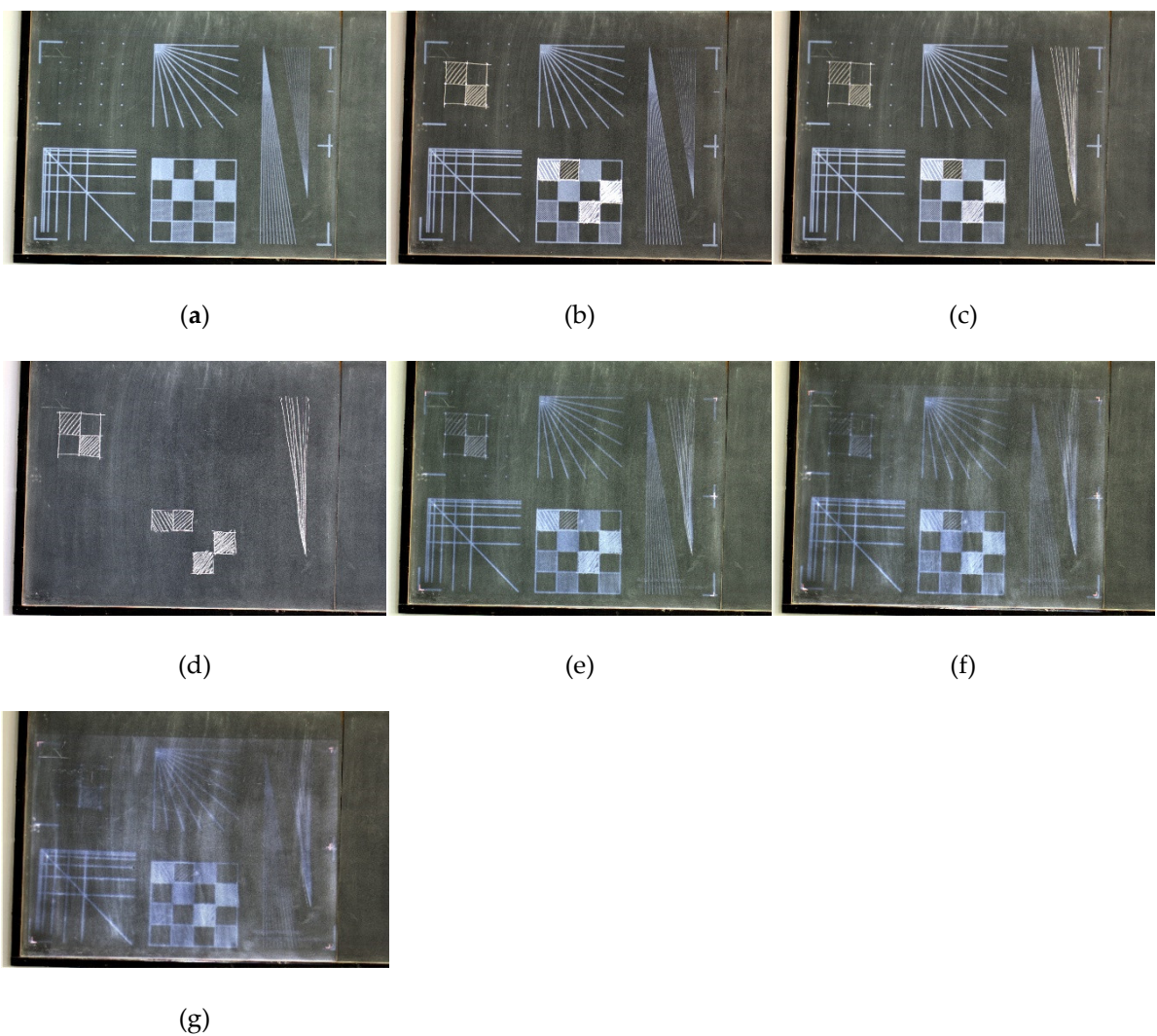

**Figure 2.** Sample JPEG images (cropped for visibility) of the projection–capture–reprojection process repeated six times over a mix of white chalk and computer prepared drawings. (**a**) First projected image used as reference; (**b**) Second reprojected image plus a chalk drawing in the upper left corner; (**c**) Third reprojected image plus chalk drawing on the right side (**d**) Only chalk drawing without overlapping projected image (**e**) Fourth reprojected image (**f**) Fifth reprojected image (**g**) Sixth reprojected image.

The measured SSIM values for the Figure 2b to 2g having as reference the Figure 2a are presented below on a scale from 0 to 1, 0 meaning no similarity and 1 meaning 100% similarity:

- Figure 2b compared with 2a has similarity of 0.6123,
- Figure 2c compared with 2a has similarity of 0.7541,
- Figure 2e compared with 2a has similarity of 0.6161,
- Figure 2f compared with 2a has similarity of 0.5942,
- Figure 2g compared with 2a has similarity of 0.5229.

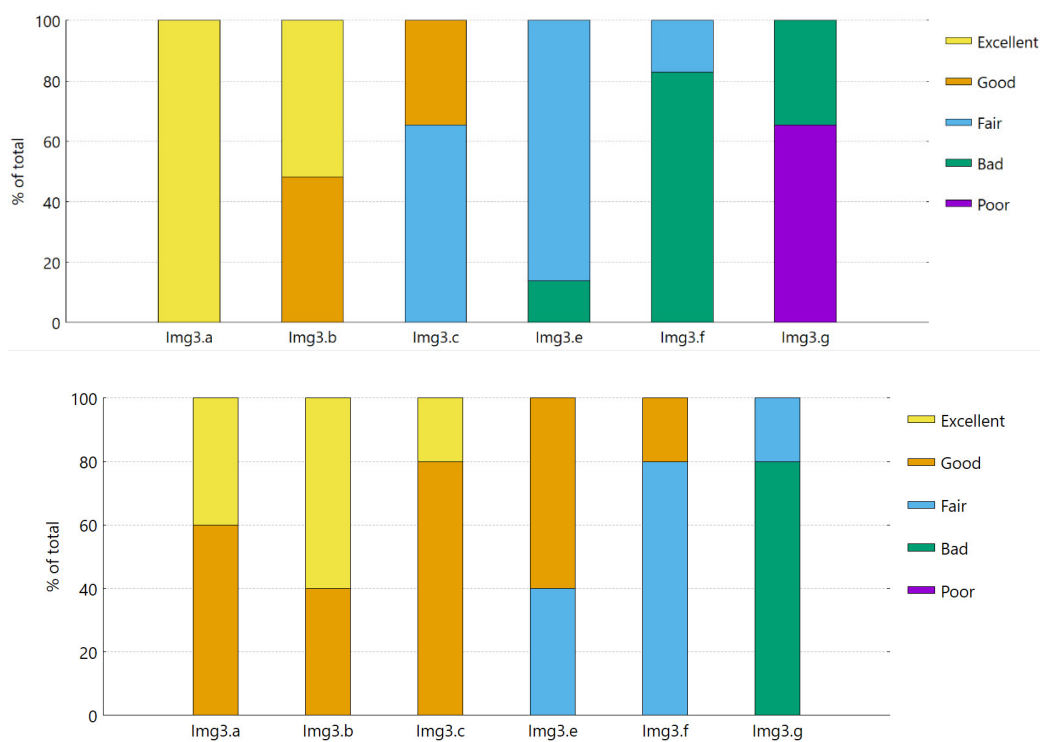

**Figure 3.** (**a**) Students' perceived quality of the images presented in Figure 2. (**b**) Teachers' perceived quality of the images presented in Figure 2.

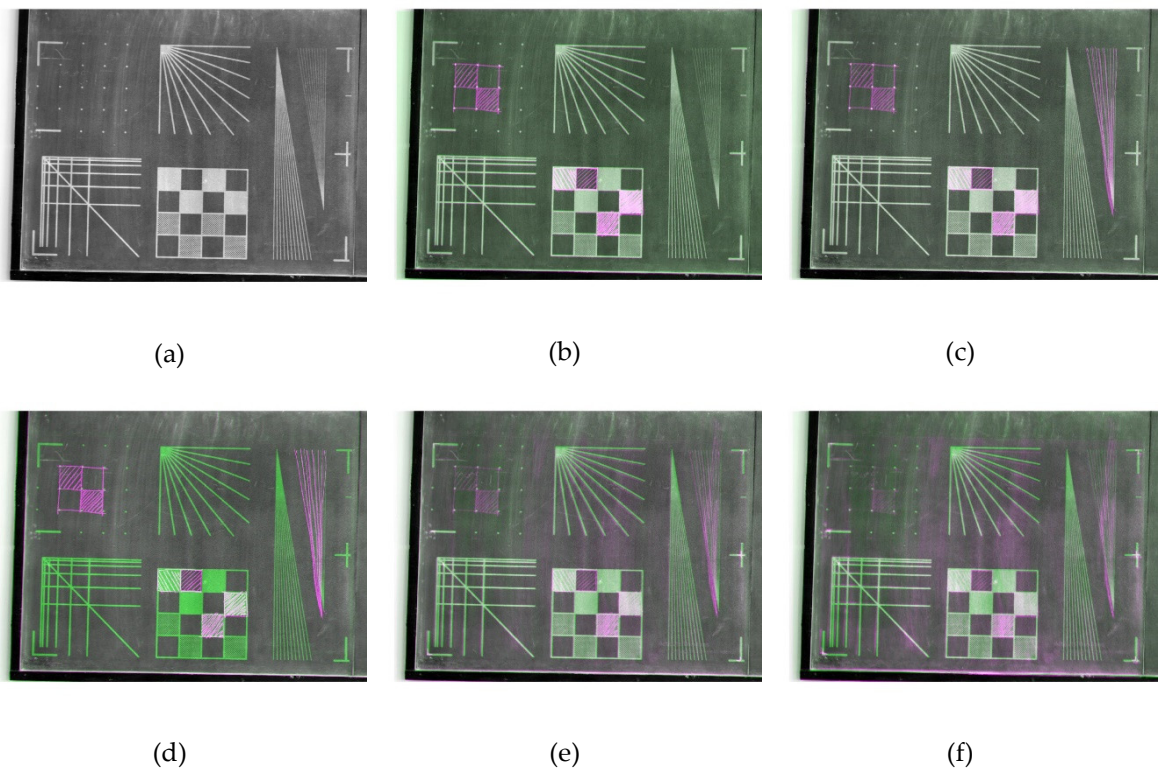

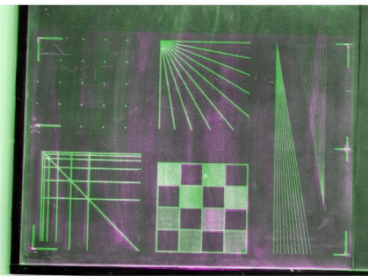

(g)

**Figure 4.** Composite RGB image composed of overlaid reference image and reprojected images. (**a**) Figure 2a overlaid with itself; (**b**) Figure 2b overlaid with 2a; (**c**) Figure 2c overlaid with 2a; (**d**) Figure 2d overlaid with 2a; (**e**) Figure 2e overlaid with 2a; (**f**) Figure 2f overlaid with 2a; (**g**) Figure 2g overlaid with 2a.

In the case of the above design references, we observe the drawing in chalk will always be more visible. When drawing in chalk on a gray surface (blackboard) over the projected reference image, the area covered by chalk will have another index of reflection and as a result the drawing has a great contrast. However, it can be observed that the contrast diminishes in areas covered exclusively by chalk drawing, without the presence of image reference projection.

Figure 2a–c represents the result of overlaying the projected reference with the chalk drawings and should not be considered reprojected images. Only Figure 2e–g are reprojections of the Figure 2c. Observing the SSIM values regarding the first reference Figure 2a we conclude that the drop in SSIM calculated value is not significant, even both teachers and students consider the third projection (Figure 2g) as almost unacceptable.

*4.2. Analyzing the Use of Projected References in Complex Design*

In the current case study, we present a concrete example of using references as support to creating complex hand drawings, as well as capturing and successively reprojecting references. In Figure 5 we depict a series of projected–captured–altered–reprojected images. In Figure 6, we present further the results of evaluating subjectively the quality of the reprojected images by questioning both teachers and students.

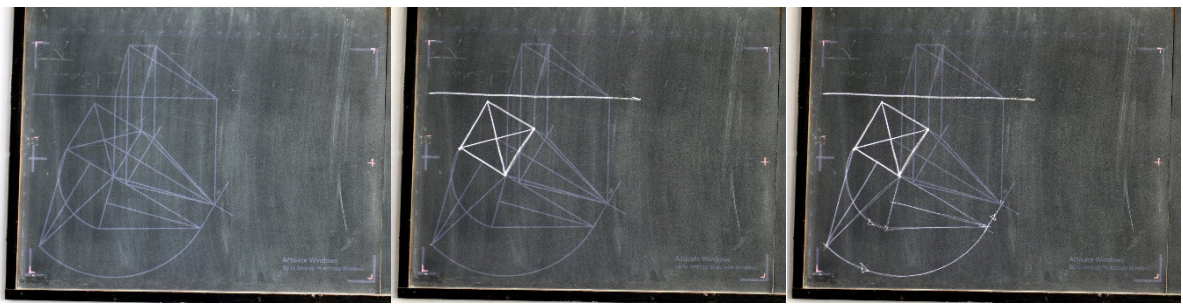

(a)　　　　　　　　　　　　(b)　　　　　　　　　　　　(c)

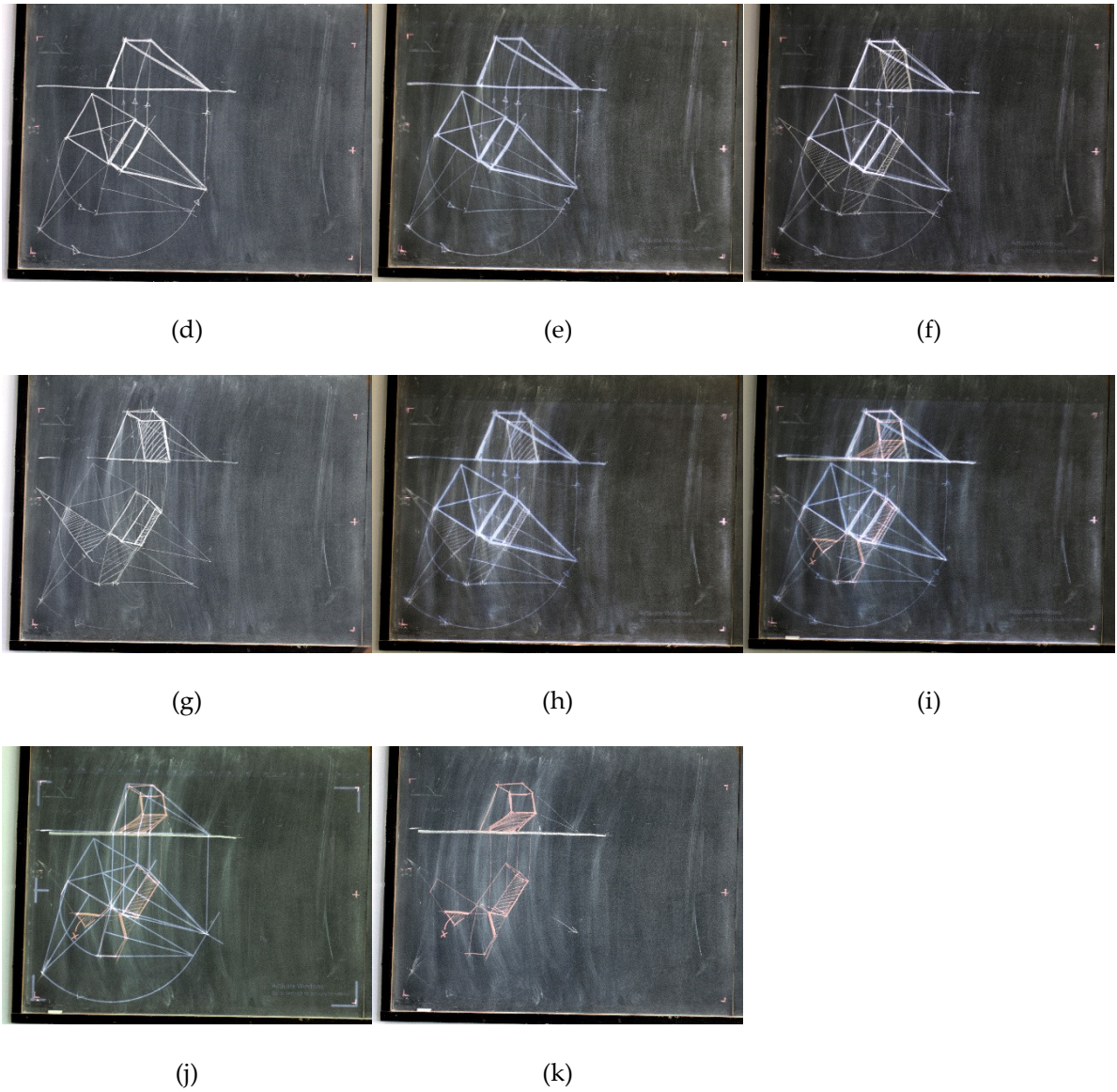

**Figure 5.** Sample JPEG images of the projection–capture–reprojection process repeated 11 times over a mix of white and colored chalk and computer prepared drawings.

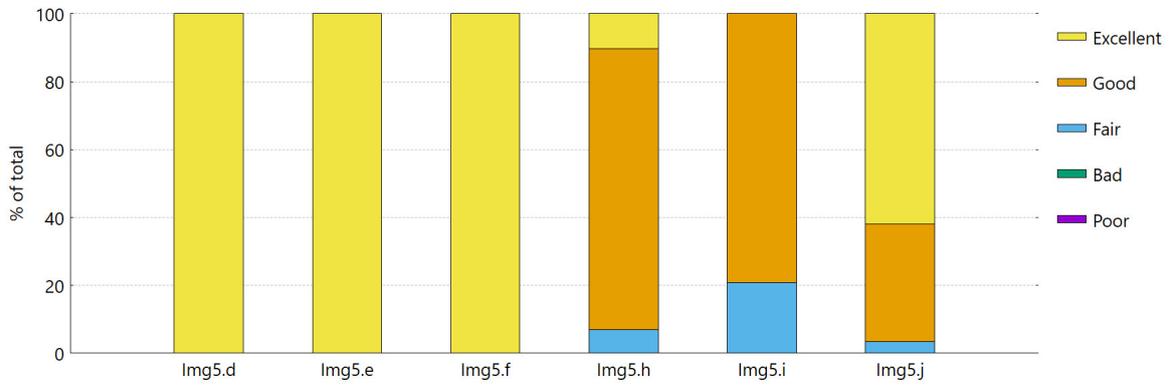

(a)

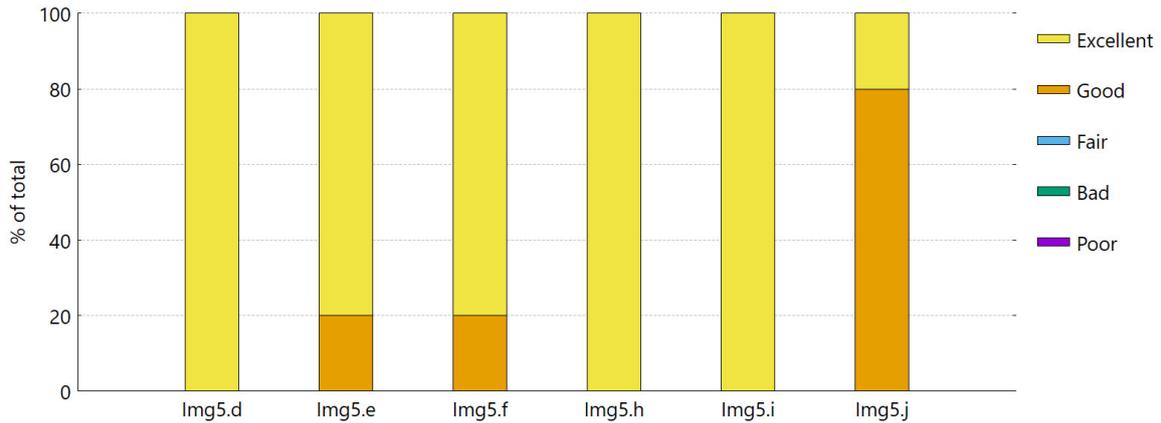

(b)

**Figure 6.** Human perceived quality of the images presented in Figure 5. (**a**) Students opinion; (**b**) Teachers opinion.

In this case study we simulated the interactive teaching process where the teacher is asked to return to previous drawings for further clarifications. Analyzing the teachers' and the students' perceptions presented in Figure 6, we observe that the process is feasible even for eleven slides, the image quality being enough to allow the drawing to be completed. The differences between the manual drawing of the last step and the initial reference are noticeable, but not high enough to make the drawing unintelligible.

*4.3. Analyzing the Quality of Successive Reprojection*

In the current case study, we analyze how the image quality is affected by successive capture and reprojection starting from an image composed of a projected computer design reference and a chalk drawing without any chalk interaction. In Figure 7 we depict a series of projected–captured–reprojected images, we evaluate objectively the similarity of each of the images in relation to the first projected one by calculating the SSIM value, and in Figure 8 we present the results obtained by questioning teachers and students. In Figure 9 we present the visual dissimilarities between images in different color bands.

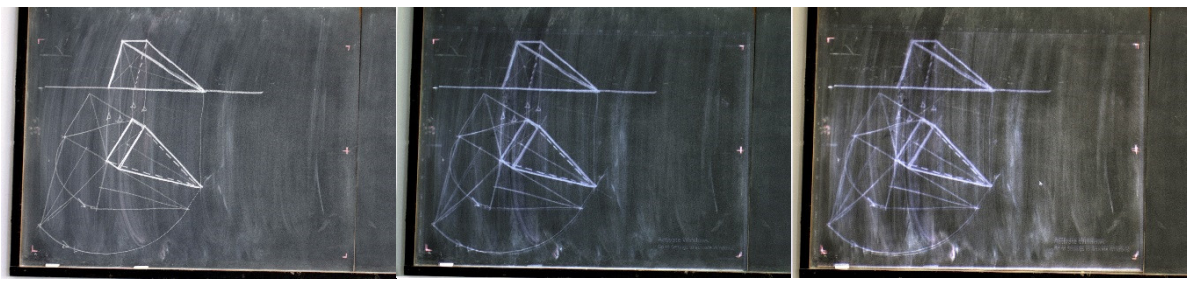

(a)　　　　　　　　　　　(b)　　　　　　　　　　　(c)

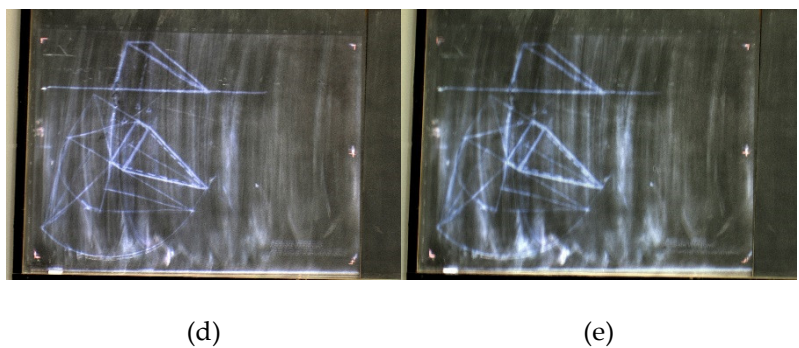

(**d**)    (**e**)

**Figure 7.** Sample JPEG images of the projection-capture-reprojection process repeated five times over a mix of white chalk and computer prepared drawings. (**a**) Mix of projection and chalk drawing; (**b**) Capture and reprojection of Figure 7a; (**c**) Capture and reprojection of Figure 7b; (**d**) Capture and reprojection of Figure 7c; (**e**) Capture and reprojection of Figure 7d.

The measured SSIM values for Figure 7b–e having as reference Figure 7a are presented below on a scale from 0 to 1, 0 meaning no similarity and 1 meaning 100% similarity:

- Figure 7b compared with 7a has similarity of 0.6555
- Figure 7c compared with 7a has similarity of 0.4593,
- Figure 7d compared with 7a has similarity of 0.4429,
- Figure 7e compared with 7a has similarity of 0.4242.

Similar with case study number one, we observe that the quality of the reprojected image deteriorates fast, a high contrast between black and white tones being observed. The human opinion indicates that from third to fourth reprojection the quality of the image is almost unacceptable. Yet, there are few situations where an image is needed to be reprojected more than twice and the teacher needs the go back and forth into presentation interacting with digital images. Observing the SSIM values regarding the first reference Figure 7a we conclude that the drop in SSIM calculated value is significant and supports the subjective human opinion.

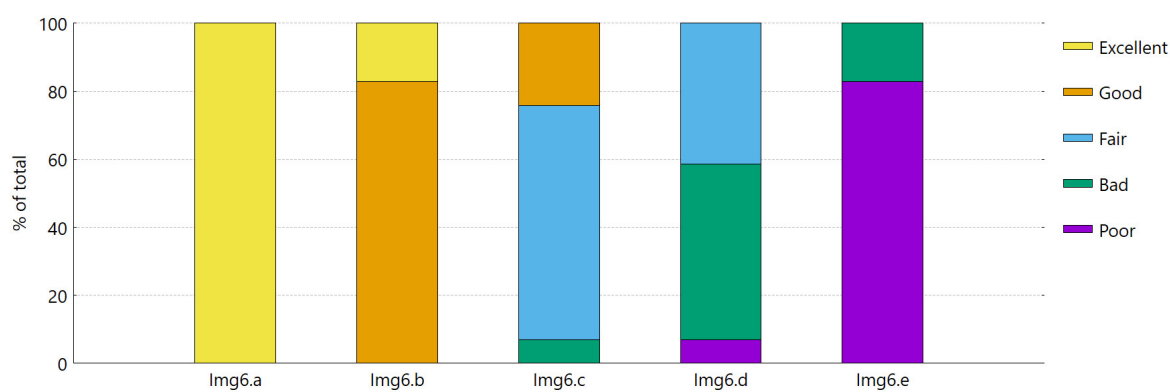

(**a**)

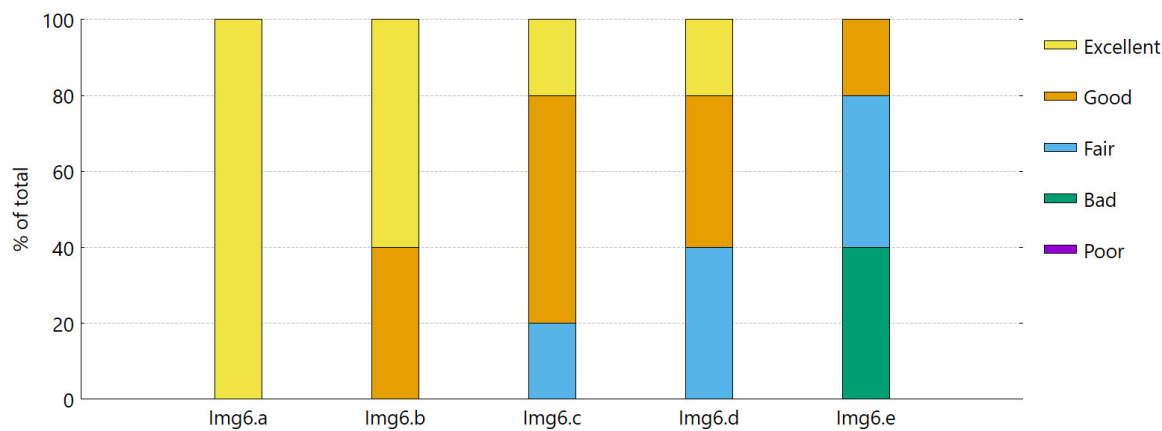

(b)

**Figure 8.** (**a**) Students' perceived quality of the images presented in Figure 7. (**b**) Teachers' perceived quality of the images presented in Figure 7.

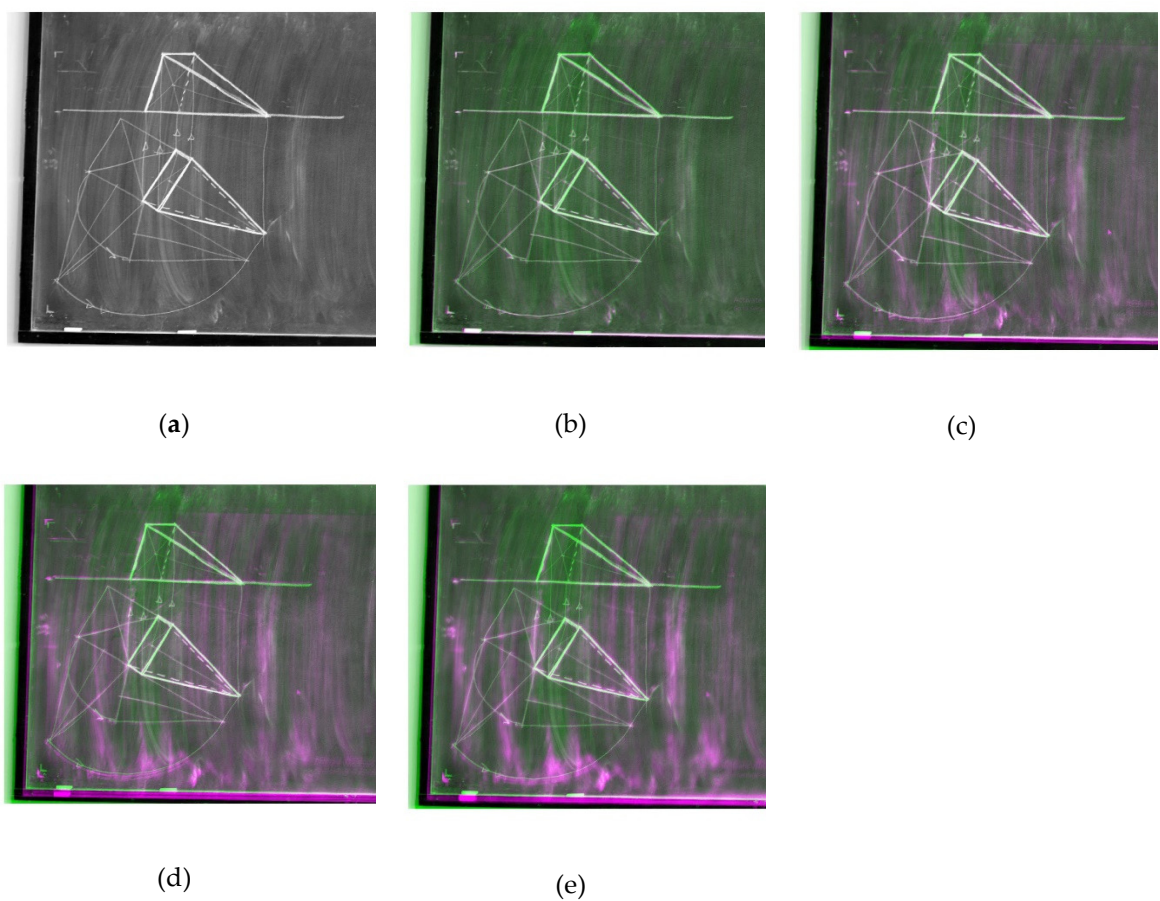

**Figure 9.** Composite RGB image composed of overlaid reference image and reprojected images. (**a**) Figure 7a overlaid over itself; (**b**) Figure 7b overlaid over 7a; (**c**) Figure 7c overlaid over 7a; (**d**) Figure 7d overlaid over 7a; (**e**) Figure 7e overlaid over 7a.

*4.4. Analyzing the Loss of Quality in Different Types of Projections*

In the current case study, we assess three image type degradation used as a reference: digital drawing having high contrast of black and white, black and white picture, and colored picture. In Figure 10 we depict a series of projected–captured–reprojected images. We evaluate objectively the similarity of each of the images in relation to the first one by calculating the SSIM value, and in Figure 11 we present the results obtained by questioning teachers and students. In Figure 12 the visual dissimilarities between images in different color bands are presented.

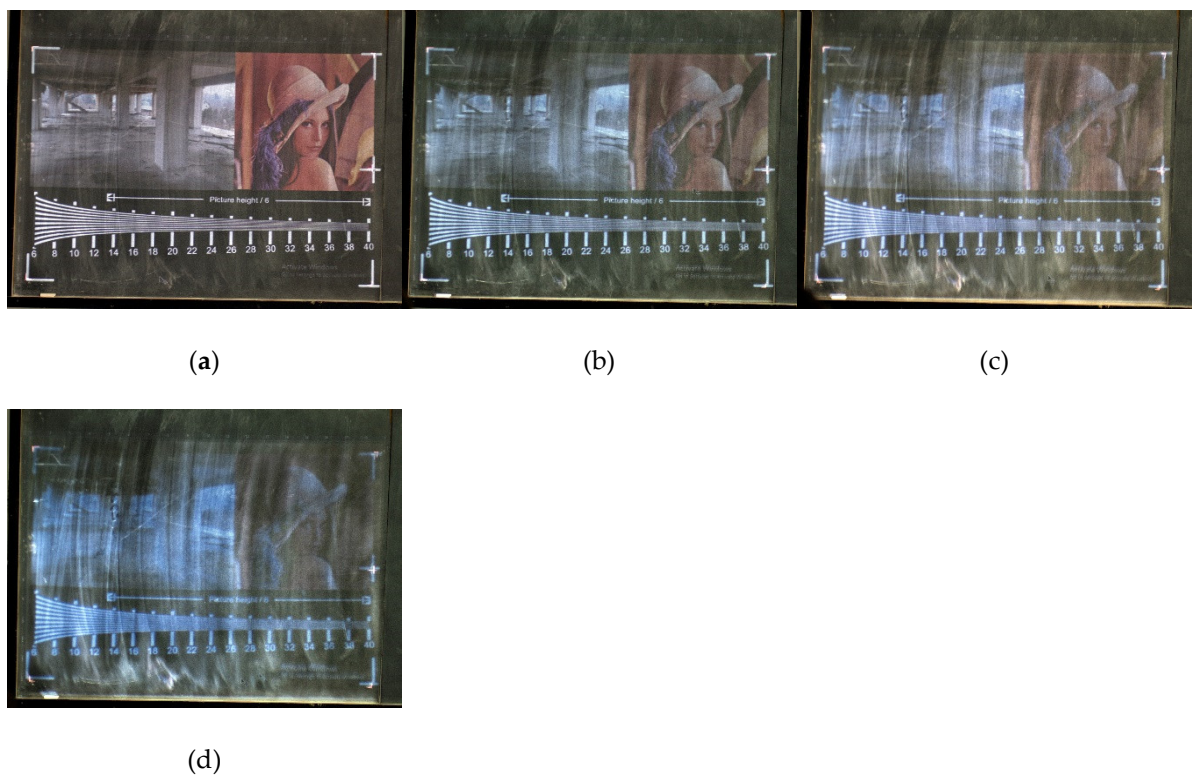

(**a**)    (**b**)    (**c**)

(**d**)

**Figure 10.** Sample JPEG images of the projection–capture–reprojection process repeated four times over a mix of computer prepared images. (**a**) First projected image used as reference; (**b**) Second reprojected image; (**c**) Third reprojected image (**d**) Fourth reprojected image.

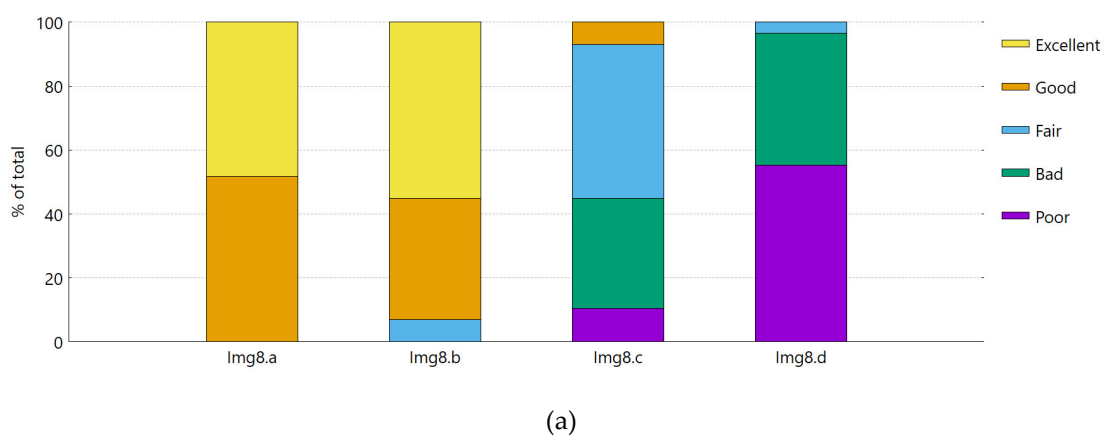

(**a**)

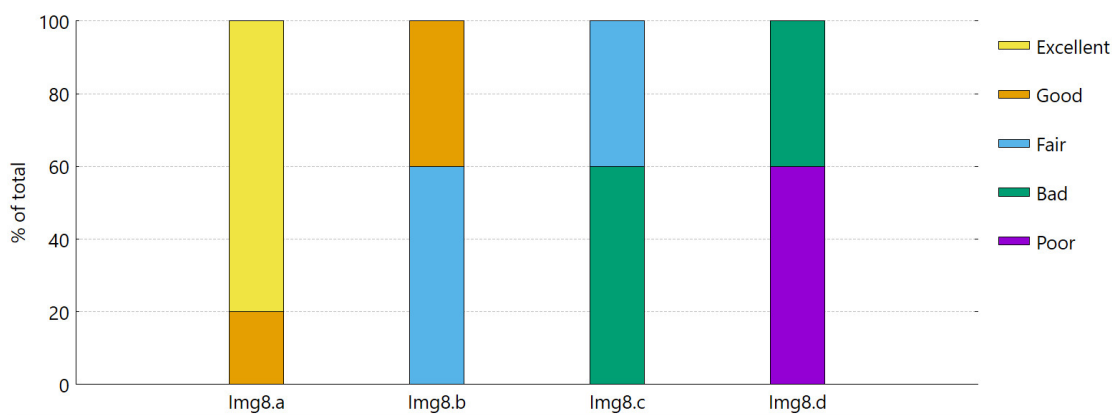

(b)

**Figure 11.** Human perceived quality of the images presented in Figure 10. (**a**) Students' opinion; (**b**) Teachers' opinion.

The measured SSIM values for Figure 10b–d having as reference Figure 10a are presented below on a scale from 0 to 1, 0 meaning no similarity and 1 meaning 100% similarity:

- Figure 10b compared with 10a has similarity of 0.6061,
- Figure 10c compared with 10a has similarity of 0.5922,
- Figure 10d compared with 10a has similarity of 0.4243.

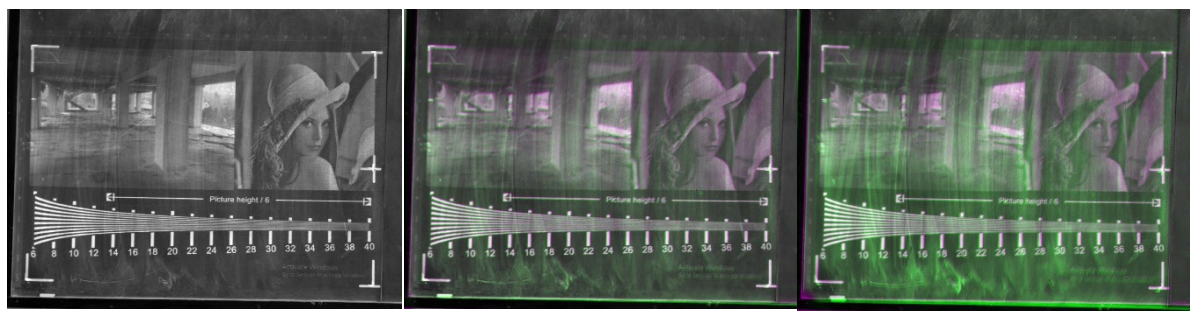

(**a**)   (**b**)   (**c**)

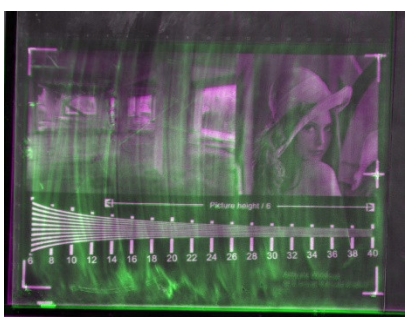

(**d**)

**Figure 12.** Composite RGB image composed of overlaid reference image and reprojected images (**a**) Figure 10a overlaid over itself; (**b**) Figure 10b overlaid over 10a; (**c**) Figure 10c overlaid over 10a; (**d**) Figure 10d overlaid over 10a.

We observe that degradation during successive reprojection is much faster for photos, with no significant difference between black and white or color, the degradation being influenced by image contrast. To maintain acceptable brightness, it is necessary to correct the contrast and brightness of the images before redesigning, emphasizing the texture of the board used as a support for the design. Yet, the drop of image quality is not so steep, and the system is still feasible albeit using a low-performance digital projector.

## 5. Discussion and Conclusion

The current study continued our previous research on hybrid design tools analyzing the limitations of the developed hybrid system regarding the perceived quality of the images being repeatedly captured, annotated, and reprojected onto the board. We validated the hybrid system by evaluating the quality of the projected and reprojected images over a blackboard, using both objective measurements and subjective human perception. Hybrid techniques of work are natural and spontaneous and make use of the most efficient available means. Even if this seems to be a logical approach, it is often rejected as a valid method due to prejudices of some students who perceive these works as an acceptance of limitations in mastering digital media, and due to prejudices of evaluators who often prefer images either in analog or in digital media. We consider hybrid design tools the efficient solution as a mixture of digital and analogic means.

The case studies were developed in order to analyze how the image quality is affected by successive capture and reprojection using the proposed hybrid system [12]. To emulate real case scenarios, the case studies were elaborated in one of the classrooms of the Architecture Department within the university, performing the setup of the hybrid system as for a real design lecture. As discussed in the previous sections, making a presentation using a digital augmented board can provide on one side adaptable and interactive answers from teachers to students' questions, and on the other side favors the bringing of multisensory perception (tactile plus auditive feedback [27] and gesture [16]) into the conversation.

We observed that the calculated SSIM values as objective assessments of the quality of reprojected images for each of the case studies are in harmony with the subjective human opinion, the increase in the reprojection number being directly proportional with the drop of SSIM value. Yet, even we consider the SSIM a supporting quality indicator of the subjective human opinion, we conjecture the human opinion is more relevant in the current context than the objective measurement. Regarding the human subjective opinion on the quality of the reprojected images we observed that the teachers group appreciated in general the same image with better score than the group of students. We think that this situation is normal because the teachers know the result and scope of the projection at a better level than the students whom at the time of assessment are also involved in an understanding and learning process. Based on the results obtained from questioning the both groups of teachers and students we conclude that the quality of the images and the references up to three consecutive reprojections over the board could be considered acceptable. We observed that the quality of the reprojected images was influenced both by the complexity of the drawing and by the perceived contrast. The proposed digital augmented board solution was also very welcomed among the teachers of geometry and design subjects because projected references, as depicted in previous sections, are very valuable in complex drawings where hand drawing alone without a reference is prone to failure. We consider the proposed augmented digital board shortens drastically the time spent by students on drafting, facilitates easy molding into design software, and creates a framework for creativity, interactivity, and quick response solutions often required within the jobsite.

Based on the results achieved in the current research, we believe that the proposed hybrid system could provide qualitative support for disciplines where complex drawings and interaction is required. Furthermore, we consider the proposed blackboard augmented system able to overcome the shortcomings of IWBs in interactive lectures, presentations, and workshops.